\documentclass[twocolumn,showpacs,showkeys,amsmath,amssymb]{revtex4}

\usepackage{graphicx}

\def\eV{\hbox{ eV}}
\def\GeV{\hbox{ GeV}}

\begin{document}

\title{Transition magnetic moments of Majorana neutrinos in supersymmetry
  without R-parity in light of neutrino oscillations}
 
\author{Marek G\'o\'zd\'z} 
\email{mgozdz@kft.umcs.lublin.pl}
\author{Wies{\l}aw A. Kami\'nski}
\email{kaminski@neuron.umcs.lublin.pl} 
\affiliation{Department of Informatics, Maria
  Curie--Sk{\l}odowska University, \\ 
  pl. Marii Curie--Sk{\l}odowskiej 5, 20-031 Lublin, Poland}
\author{Fedor \v Simkovic} 
\email{fedor.simkovic@fmph.uniba.sk}
\altaffiliation{On leave of absence from Department
  of Nuclear Physics, Comenius University, Mlynsk\'a Dolina F1, SK--842
  15 Bratislava, Slovakia.}
\author{Amand Faessler} 
\email{amand.faessler@uni-tuebingen.de}
\affiliation{Institute f\"ur Theoretische Physik der Universit\"at
  T\"ubingen, D-72076 T\"ubingen, Germany}

\begin{abstract}
  The transition magnetic moments of Majorana neutrinos $\mu_{\nu_{ij}}$
  ($ij=e\mu, ~e\tau,~\mu\tau$) are calculated in grand unified theory
  (GUT) constrained Minimal Supersymmetric Standard Model (MSSM) with
  explicit $R$-parity violation. It is assumed that neutrinos acquire
  masses via one-loop (quark-squark and lepton-slepton) radiative
  corrections. The mixing of squarks, sleptons, and quarks is considered
  explicitly. The connection between $\mu_{\nu_{ij}}$ and the entries of
  neutrino mass matrix is studied. The current upper limits on
  $\mu_{\nu_{ij}}$ are deduced from the elements of phenomenological
  neutrino mass matrix, which is reconstructed using the neutrino
  oscillation data and the lower bound on the neutrinoless double beta
  decay half-life.  Further, the results for $\mu_{\nu_{e\mu}}$,
  $\mu_{\nu_{e\tau}}$ and $\mu_{\nu_{\mu\tau}}$ are presented for the
  cases of inverted and normal hierarchy of neutrino masses and
  different SUSY scenarios. The largest values are of the order of
  $10^{-17}$ in units of Bohr magneton.
\end{abstract}


\maketitle 

\section{Introduction}

Supersymmetric extensions of the Standard Model (SM) give natural
framework for solving the hierarchy problem. They lead also to
unification of interactions at energies of the order of $m_{GUT} \sim
10^{16} \GeV$. Supersymmetry (SUSY) itself is required by string theory,
the best candidate by now which resolves most of the SM problems.

The Minimal Supersymmetric Standard Model (MSSM) is the minimal in the
interaction and particle content consistent extension of the SM. It
provides rich phenomenology by the introduction of twice as many
particles as we know from the SM, the so-called superpartners of usual
particles. It possesses also an accidental symmetry called $R$-parity,
defined by means of the baryon ($B$), lepton ($L$), and spin ($S$)
numbers as $R=(-1)^{3B+L+2S}$. Conservation of $R$-parity implies that
SUSY particles cannot decay into non-SUSY ones. As a consequence the
lightest SUSY particle must be stable and may be considered as a good
candidate for dark matter. Theoretically, however, nothing motivates
$R$-parity conservation and many models which allow for breaking of this
symmetry have been considered. Among them one can distinguish three
groups. In the first one $R$-parity violation (RpV) is introduced as a
spontaneous process triggered by a non-zero vacuum expectation value of
some scalar field \cite{aul83}. Other possibilities include the
introduction of additional bi- \cite{valle} or trilinear
\cite{rbreaking} RpV terms in the superpotential.

The most pressing motivation for looking on physics beyond the SM comes
from the discovery of neutrino oscillations \cite{SK,SNO,Kamland,K2K}.
The experimental evidence for this process clearly indicates that
neutrinos do have non-zero masses. What is more, the flavour and mass
eigenstates are not the same, which leads to mixing between them. In
experiments focused on neutrino oscillations one can measure the mixing
angles as well as the differences of masses squared. For absolute values
of masses one has to study the large scale structures of the Universe,
the endpoint of the electron spectrum in beta decay of Tritium
\cite{tritium}, or search for signals of the hypothetical neutrinoless
double beta decay ($0\nu\beta\beta$). The latter is also the only known
process which distinguishes between Majorana and Dirac-like neutrinos.
Its observation will prove the Majorana nature of these particles.

The importance of theoretical studies of various properties of neutrinos
is obvious. In the present paper we focus on the transition magnetic
moment, which is generated by interactions of quark--squark and
lepton--slepton self-energy loops with an external photon. This work is
a natural continuation and extension of calculations presented in
Refs.~\cite{Haug,Bhatta,Abada,mg-art6}.

In the following section we define the model, describe in detail our
method of obtaining the supersymmetric particle spectrum, and the way in
which unification (GUT constraints) is imposed. In Sec. III we construct
the neutrino mass matrices in loop mechanism, as well as calculate the
transition magnetic moments. In our present approach the squark and
quark mixings are taken into account. We present our results and state
the conclusions in the last part of the paper.

\section{GUT constraints and particle spectrum}

Let us fix our framework to be the Minimal Supersymmetric Standard Model
with supersymmetry breaking mediated by the gravity force (SUGRA MSSM).
We follow the conventions and notation from Ref.~\cite{HirschValle}. We
allow for $R$-parity violation by considering explicitly bilinear and
trilinear RpV terms in the superpotential. We focus, however, on the
trilinear part, as the effects we want to study come from 1-loop
processes, while the bilinear terms induce neutrino masses at tree
level.

The $R$-parity conserving part of the superpotential of MSSM has the
form
\begin{eqnarray}
  W^{MSSM} &=& \epsilon_{ab} [(\mathbf{Y}_E)_{ij} L_i^a H_1^b \bar E_j
    + (\mathbf{Y}_D)_{ij} Q_i^{ax} H_1^b \bar D_{jx} \nonumber \\
    &+& (\mathbf{Y}_U)_{ij} Q_i^{ax} H_2^b \bar U_{jx} + \mu H_1^a H_2^b ],
\label{wmssm}
\end{eqnarray}
while its RpV part reads
\begin{eqnarray}
  W^{RpV} &=& \epsilon_{ab}\left[
    \frac{1}{2} \lambda_{ijk} L_i^a L_j^b \bar E_k
    + \lambda'_{ijk} L_i^a Q_j^{xb} \bar D_{kx} \right] \nonumber \\
  &+& \frac{1}{2}\epsilon_{xyz} \lambda''_{ijk}\bar U_i^x\bar
  D_j^y \bar D_k^z + \epsilon_{ab}\kappa^i L_i^a H_2^b.
\end{eqnarray}
The {\bf Y}'s are 3$\times$3 Yukawa matrices. $L$ and $Q$ are the $SU(2)$
left-handed doublets while $\bar E$, $\bar U$ and $\bar D$ denote the
right-handed lepton, up-quark and down-quark $SU(2)$ singlets,
respectively. $H_1$ and $H_2$ mean two Higgs doublets. We have introduced
color indices $x,y,z = 1,2,3$, generation indices $i,j,k=1,2,3$ and the SU(2)
spinor indices $a,b,c = 1,2$. 

The introduction of RpV terms implies the existence of lepton or barion
number violating processes, like the unobserved proton
decay. Fortunatelly one may keep only one type of terms and it is not
necessary to have both at one time. In order to get rid of too rapid
proton decay and to allow for lepton number violating processes, like
the neutrinoless double beta decay, it is customary to set
$\lambda''=0$.

We supply the model with scalar mass term
\begin{eqnarray}
{\cal L}^{mass} &=& \mathbf{m}^2_{H_1} h_1^\dagger h_1 +
                    \mathbf{m}^2_{H_2} h_2^\dagger h_2 +
     q^\dagger \mathbf {m}^2_Q q + l^\dagger \mathbf {m}^2_L l \nonumber \\
&+&  u \mathbf {m}^2_U u^\dagger + d \mathbf {m}^2_D d^\dagger +
     e \mathbf {m}^2_E e^\dagger,
\end{eqnarray}
soft gauginos mass term ($\alpha=1,...,8$ for gluinos)
\begin{equation}
  {\cal L}^{gaug.} = \frac12 \left( 
  M_1 \tilde{B}^\dagger \tilde{B} + 
  M_2 \tilde{W_i}^\dagger \tilde{W^i} +
  M_3 \tilde{g_\alpha}^\dagger \tilde{g^\alpha} + h.c.\right ),
\end{equation}
as well as the supergravity mechanism of supersymmetry breaking, by
introducing the Lagrangian
\begin{eqnarray}
  {\cal L}^{soft} &=& \epsilon_{ab} [(\mathbf{A}_E)_{ij} l_i^a h_1^b \bar e_j
    + (\mathbf{A}_D)_{ij} q_i^{ax} h_1^b \bar d_{jx} \nonumber \\
    &+& (\mathbf{A}_U)_{ij} q_i^{ax} h_2^b \bar u_{jx} + B \mu h_1^a h_2^b +
    B_2 \epsilon_i l_i^a h_2^b],
\end{eqnarray}
where lowercase letters stand for scalar components of respective chiral
superfields, and 3$\times$3 matrices {\bf A} as well as $B\mu$ and $B_2$ are
the soft breaking coupling constants.

The procedure of finding a GUT constrained low energy spectrum of the
model consists of a few steps.

We take into account mass thresholds where SUSY particles start to
contribute \cite{kanekolda} and use the SM 1--loop renormalization group
equations (RGE) \cite{drtjones} below appropriate threshold and 1--loop
MSSM RGE \cite{MartinVaughn} above it. The 2--loop corrections as well
as corrections coming from the presence of RpV couplings are small and
do not change the results by more than few percent \cite{mg-art1}.

Initially all the thresholds are set to 1 TeV and are dynamically
modified during the running of mass parameters. The values of Yukawa
couplings {\bf Y} at $m_Z$ are given by lepton and quark mass matrices
$M$ \cite{ChoMisiak}
\begin{eqnarray}
  M_U &=& v_u \mathbf{S}_{U_R} \mathbf{Y}_U^T
  \mathbf{S}_{U_L}^\dagger, \nonumber \\
  M_D &=& v_d \mathbf{S}_{D_R} \mathbf{Y}_D^T
  \mathbf{S}_{D_L}^\dagger, \\
  M_E &=& v_d \mathbf{S}_{E_R} \mathbf{Y}_E^T
  \mathbf{S}_{E_L}^\dagger, \nonumber
\end{eqnarray}
where $v_d=\langle H_1^0 \rangle$ and $v_u=\langle H_2^0 \rangle$ are
the neutral Higgs vacuum expectation values. Their ratio defines the
angle $\beta$ through the relation $v_u/v_d=\tan\beta$. {\bf S} matrices
perform diagonalization so that one obtains eigenstates in the mass
representation.

After evolving the dimensionless couplings from the electroweak scale
$m_Z$ up to $m_{GUT} = 2 \times 10^{16}$ GeV, we unify the masses of
gauginos, sfermions and squarks to be equal to a common mass parameter
$m_0$. We set also the masses of all fermions to a common value
$m_{1/2}$. The trilinear soft couplings at $m_{GUT}$ are set according
to the formula
$$
\mathbf{A}_i = A_0 \mathbf{Y}_i,
$$
with $A_0$ being another input parameter. We construct the squark,
slepton, chargino and neutralino mass matrices and in the next step
evolve all the quantities down to $m_Z$. During that running the
tree-level Higgs potential is minimized, ie. the following set of
equations is solved:
\begin{eqnarray}
  (\mu B)^2 &>& ( \left|\mu\right|^2 + m_{H_1}^2 )
                ( \left|\mu\right|^2 + m_{H_2}^2 ), \nonumber \\
  2B\mu &<& 2 \left|\mu\right|^2 + m_{H_1}^2 + m_{H_2}^2.
\label{EWSB}
\end{eqnarray}
In fact one should minimize the full 1--loop Higgs potential, but as the
first approximation we consider only the tree-level. We compensate this
by adding radiative corrections, which contain functions of particle
mass eigenstates generated by elektroweak symmetry breaking (EWSB)
mixing.  In that way a proper EWSB mechanism is included in our
procedure.

To obtain the physical masses of supersymmetric particles one has to
diagonalize proper mass matrices. For the down squarks we have
%
\begin{widetext}
\begin{equation}
  m_{\tilde d^k}^2 =
  \begin{pmatrix}
    m_{\tilde d_L^k}^2 + m_{d^k}^2 - \frac16(2m_W^2+m_Z^2)\cos2\beta &
    -m_{d^k} ((\mathbf{A}_D)_{kk} + \mu\tan\beta) \cr
    -m_{d^k} ((\mathbf{A}_D)_{kk} + \mu\tan\beta) &
    m_{\tilde d_R^k}^2 + m_{d^k}^2 + \frac13(m_W^2-m_Z^2)\cos2\beta
  \end{pmatrix},
\end{equation}
and for sleptons:
\begin{equation}
  m_{\tilde e^k}^2 =
  \begin{pmatrix}
    m_{\tilde e_L^k}^2 + m_{e^k}^2 - \frac12(2m_W^2-m_Z^2)\cos2\beta &
    -m_{e^k} ((\mathbf{A}_E)_{kk} + \mu\tan\beta) \cr
    -m_{e^k} ((\mathbf{A}_E)_{kk} + \mu\tan\beta) &
    m_{\tilde e_R^k}^2 + m_{e^k}^2 + (m_W^2-m_Z^2)\cos2\beta
  \end{pmatrix},
\end{equation}
\end{widetext}
%
where $m_W^2 = m_Z^2 \cos^2\theta_W$, and the Weinberg weak mixing angle
is $\sin^2\theta_W=0.22$. We have denoted by $m_{d^k}$ and $m_{e^k}$,
$k=1,2,3$, the masses of down quarks and charged leptons respectively.
The $L$-handed elements are the eigenvalues of the running mass
parameters $\mathbf {m}^2_Q$ and $\mathbf {m}^2_L$ whereas the
$R$-handed ones are the eigenvalues of the singlet parameters $\mathbf
{m}^2_D$ and $\mathbf {m}^2_E$. They are obtained from the RGE procedure
after diagonalization. The diagonalization procedure involves
multiplication by orthogonal matrices which in the standard
trigonometric parametrization introduce mixing angles between the weak
and mass eigenstates in the following way:
\begin{eqnarray}
 \tilde d_L &=& \phantom{-}\tilde d_1 \cos\theta + \tilde d_2 \sin\theta, \cr
 \tilde d_R &=& -\tilde d_1 \sin\theta + \tilde d_2 \cos\theta,
\end{eqnarray}
\begin{eqnarray}
 \tilde e_L &=& \phantom{-}\tilde e_1 \cos\phi + \tilde e_2 \sin\phi, \cr
 \tilde e_R &=& -\tilde e_1 \sin\phi + \tilde e_2 \cos\phi.
\end{eqnarray}
For completeness we list below the explicit expressions for the mixing
angles and mass eigenstates. For the $d$-type squarks we finish with
\begin{eqnarray}
  \sin 2\theta^k &=& -2 m_{d^k} ((\mathbf{A}_D)_{kk} + \mu \tan\beta) \cr
  &\times& [(m_{\tilde d_L^k}^2 - m_{\tilde d_R^k}^2 - 0.34 m_Z^2
    \cos 2\beta)^2 \cr
  &+& 4 m_{d^k}^2 ((\mathbf{A}_D)_{kk} + \mu \tan\beta)^2]^{-1/2},
\label{mixingq}
\end{eqnarray}
$(\mathbf{A}_D)_{kk}$ being the $k$-th eigenvalue of $\mathbf{A}_D$, and
\begin{eqnarray}
  m_{\tilde d_{1,2}^k}^2 &=&
    \frac12 (m_{\tilde d_L^k}^2 + m_{\tilde d_R^k}^2) 
    - \frac14 m_Z^2 \cos 2\beta \nonumber \\
    &+& m_{d^k} \Big ( m_{d^k} \mp \frac{(\mathbf{A}_D)_{kk} + 
      \mu \tan\beta}{\sin 2\theta^k} \Big).
  \end{eqnarray}
For sleptons the analogous expressions read:
\begin{eqnarray}
  \sin 2\phi^k &=& -2 m_{e^k} ((\mathbf{A}_E)_{kk} + \mu \tan\beta) \cr
  &\times& [(m_{\tilde e_L^k}^2 - m_{\tilde e_R^k}^2 - 0.04 m_Z^2
    \cos 2\beta)^2 \cr
  &+& 4 m_{e^k}^2 ((\mathbf{A}_E)_{kk} + \mu \tan\beta)^2]^{-1/2}
\label{mixingl}
\end{eqnarray}
\begin{eqnarray}
  m_{\tilde e_{1,2}^k}^2 &=&
  \frac12 (m_{\tilde e_L^k}^2 + m_{\tilde e_R^k}^2) 
  - \frac14 m_Z^2 \cos 2\beta \nonumber \\
  &+& m_{e^k} \Big ( m_{e^k} \mp \frac{(\mathbf{A}_E)_{kk} + 
    \mu \tan\beta}{\sin 2\phi^k} \Big).
\end{eqnarray}

Each obtained solution is checked against various conditions. These are
(1) finite values of Yukawa couplings at the GUT scale; (2) requirement
of physically acceptable mass eigenvalues at low energies; (3) FCNC
phenomenology ($b \to s \gamma$ processes).

The model we have described has only five free parameters: three GUT
parameters $A_0$, $m_0$, $m_{1/2}$, the sign of $\mu$ and $\tan \beta$.

\section{Neutrino masses and transition magnetic moments}

Many different proposals of generating neutrino mass matrix may be found
in the literature \cite{nu-mass}. In the present paper we use the
loop mechanism of generating Majorana neutrino masses, possible in RpV
models \cite{Haug,Bhatta,Abada,mg-art6}. It is also well known that by
considering photon interactions with the particles in the loop a
transition magnetic moment is generated.

\subsection{The neutrino masses in R-parity breaking MSSM}

Let us first recall the known results. Within the R-parity breaking 
MSSM neutrinos are, in general, massive. 

In the lowest order, the
contribution to the neutrino mass matrix reads \cite{Haug}
\begin{eqnarray}
  && {\cal M}^{tree}_{ii'} = \Lambda_i \Lambda_{i'} \ g_2^2 \cr
  &&\times  \frac{M_1 + M_2 \tan^2\theta_W}
  {4(\mu m_W^2 (M_1 + M_2 \tan^2\theta_W)\sin2\beta - M_1 M_2\mu^2)},
  \cr &
\label{Mtree}
\end{eqnarray}
where $\Lambda_i = \mu \langle \tilde{\nu}_i \rangle - v_d \kappa_i$,
and $\langle \tilde{\nu}_i \rangle$ are the vacuum expectation values of
the sneutrino fields.

%
\begin{figure}
 \includegraphics[width=0.4\textwidth]{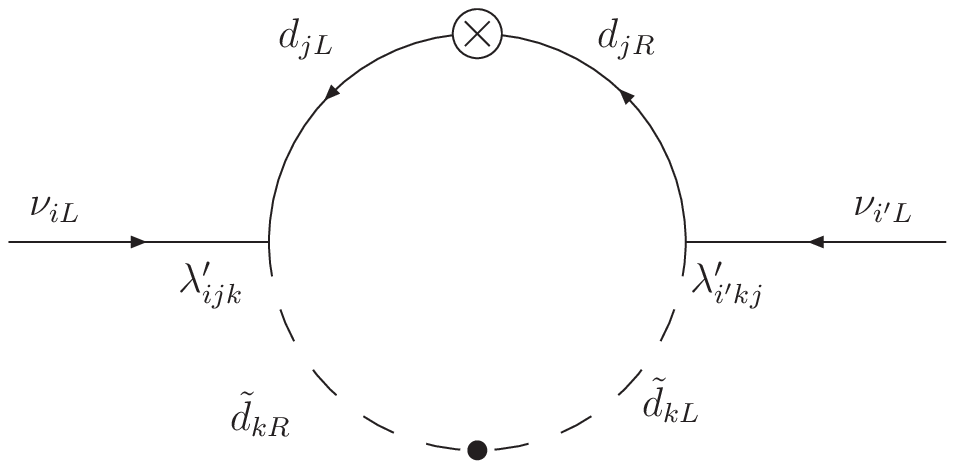}
 \includegraphics[width=0.4\textwidth]{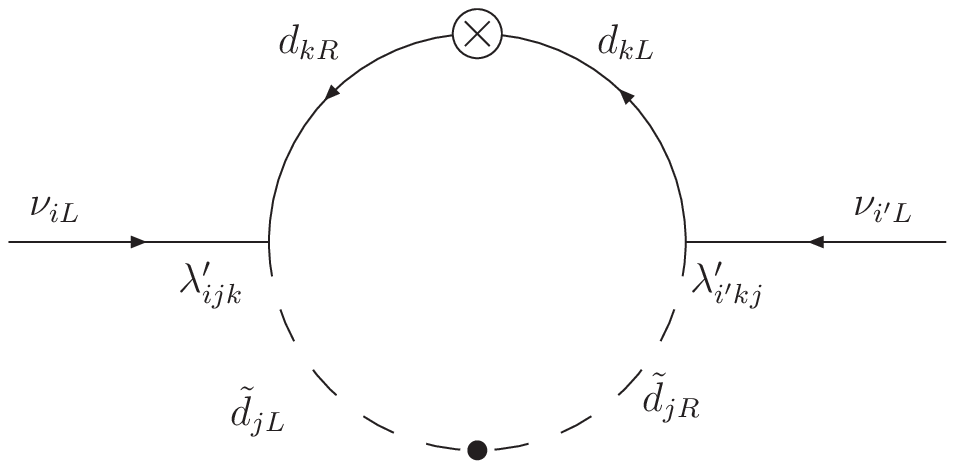}
 \caption{\label{Fig1} Feynman diagrams representing the squark-quark
 loop contribution to the Majorana neutrino mass.}
\end{figure}
\begin{figure}
 \includegraphics[width=0.4\textwidth]{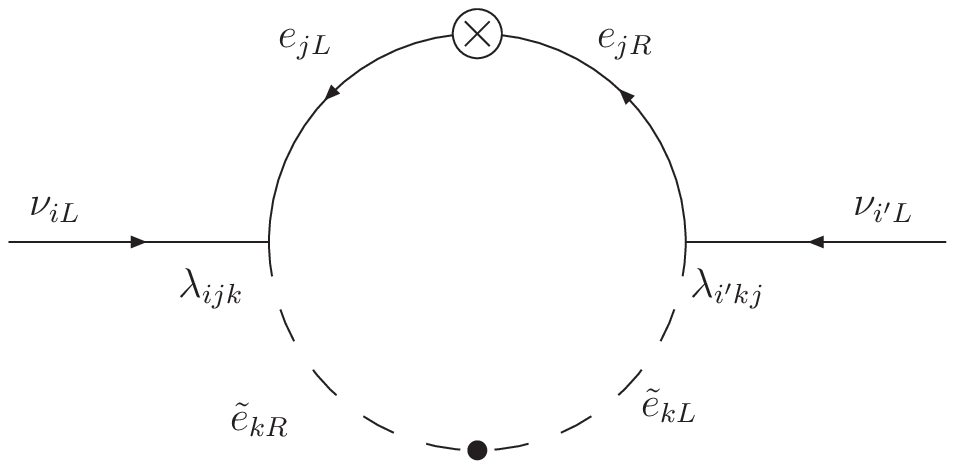}
 \includegraphics[width=0.4\textwidth]{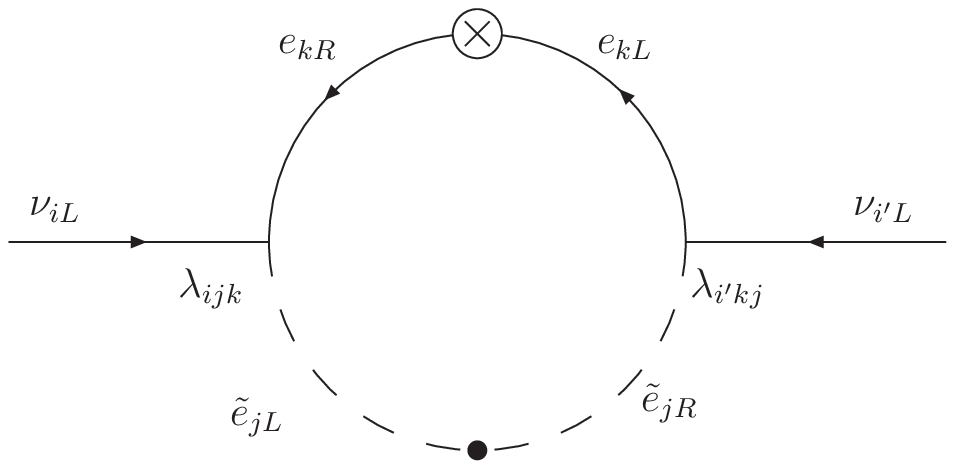}
 \caption{\label{Fig2} Feynman diagrams representing the slepton-lepton
 loop contribution to the Majorana neutrino mass.}
\end{figure}
%

Going beyond the tree level one may consider diagrams depicted on
Fig.~\ref{Fig1}. The down-squark mixing is realized through the angle
$\theta$. The quark mixing may be taken into account in two ways: as
mixing of down-type quarks and as mixing of up-type quarks (see
discussion in Ref.~\cite{dedes}). Since in our case the $d$-quarks enter
the loops, only their mixing will influence the results. The case of
$u$-quarks mixing is equivalent to switching the mixing off and has no
impact on the outcome of our calculations. We denote by $V$ the standard
Cabibbo--Kobayashi--Maskawa (CKM) quark mixing matrix.

Altogether the contribution to the neutrino mass matrix coming from the
squark-quark loop may be expressed as
\begin{eqnarray}
  {\cal M}_{ii'}^q &=&
  \frac{3}{16\pi^2} \sum_{jkl} \Bigg\{ 
  \left(
    \lambda'_{ijk}\lambda'_{i'kl} \sum_a V_{ja} V_{la} v^q_{ak} m_{d^a}
  \right ) \nonumber \\
  &+& \left (
    \lambda'_{ijk}\lambda'_{i'lj} \sum_a V_{ka} V_{la} v^q_{aj} m_{d^a}
  \right ) \Bigg \},
\label{Mqq}
\end{eqnarray}
where the loop integral is
\begin{eqnarray}
  v^q_{jk} = \frac{\sin 2\theta^k}{2} \left(
    \frac{\ln x_2^{jk}}{1-x_2^{jk}} - 
    \frac{\ln x_1^{jk}}{1-x_1^{jk}} \right ).
\label{vq}
\end{eqnarray}
In the above expressions we have introduced two dimensionless
quantities, $x_1^{jk} \equiv m_{d^j}^2 / m_{\tilde d_1^k}^2$ and
$x_2^{jk} \equiv m_{d^j}^2 / m_{\tilde d_2^k}^2$. The factor 3 in
Eq.~(\ref{Mqq}) comes from the summation over three quark colors.

The situation is simpler for the slepton-lepton loop (Fig.~\ref{Fig2}),
as the lepton mixing is much weaker and we feel justified to neglect
it. By analogy with Eq.~(\ref{Mqq}) we obtain 
\begin{equation}
  {\cal M}_{ii'}^\ell = 
  \frac{1}{16\pi^2} \sum_{jk} \lambda_{ijk}\lambda_{i'kj} 
  (v^\ell_{jk} m_{e^j} + v^\ell_{kj} m_{e^k}),
\label{Mll}
\end{equation}
with the loop integral taking the form
\begin{eqnarray}
  v^\ell_{jk} = \frac{\sin 2\phi^k}{2} \left(
    \frac{\ln y_2^{jk}}{1-y_2^{jk}} - 
    \frac{\ln y_1^{jk}}{1-y_1^{jk}} \right ),
\end{eqnarray}
where now $y_1^{jk} \equiv m_{e^j}^2 / m_{\tilde e_1^k}^2$ and
$y_2^{jk} \equiv m_{e^j}^2 / m_{\tilde e_2^k}^2$.

The functions $v^q_{jk}$ and $v^\ell_{jk}$, which are free of $R$-parity
breaking SUSY paprameters, can be calculated by running the MSSM RGE, in
which the whole low energy particle/sparticle spectrum is generated.  In
our calculation we have used the following values for the quark sector:
$m_u=5$ MeV, $m_d=9$ MeV, $m_s=175$ MeV, $m_c=1.5$ GeV, $m_b=5$ GeV,
$m_t=174$ GeV, as well as the CKM matrix in the form
\begin{equation}
  V = 
  \begin{pmatrix}
    \phantom{-}0.9755 & \phantom{-}0.2195 & 0.0031 \cr
    -0.2201 & \phantom{-}0.9746 & 0.0390 \cr
    -0.0055 & -0.0387 & 0.9989
  \end{pmatrix}.
\end{equation}
This corresponds to the sines of quark mixing angles $s_{12} = 0.2195$,
$s_{23} = 0.039$, and $s_{13} = 0.0031$ \cite{dedes} in the standard
trigonometric parametrization. We have neglected the possible appearance
of the CP violating phase in the quark sector.

The values of the calculated loop integrals have been arranged for
clarity in the form of matrices. For the input parameters $A_0=100
\GeV$, $m_0=m_{1/2}=150 \GeV$, $\tan\beta=19$, $\mu>0$ we get:
\begin{equation*}
  v^q = 
  \begin{pmatrix}
    2.361\times 10^{-4} & 4.593\times 10^{-3} & 1.687\times 10^{-1} \cr
    2.361\times 10^{-4} & 4.593\times 10^{-3} & 1.687\times 10^{-1} \cr
    2.358\times 10^{-4} & 4.587\times 10^{-3} & 1.684\times 10^{-1}
  \end{pmatrix},
\end{equation*}
\begin{equation*}
  v^\ell = 
  \begin{pmatrix}
    6.547\times 10^{-5} & 1.310\times 10^{-2} & 2.528\times 10^{-1} \cr
    6.547\times 10^{-5} & 1.310\times 10^{-2} & 2.528\times 10^{-1} \cr
    6.541\times 10^{-5} & 1.308\times 10^{-2} & 2.525\times 10^{-1}
  \end{pmatrix}.
\end{equation*}
For the second set of parameters $A_0=500 \GeV$, $m_0=m_{1/2}=1000
\GeV$, $\tan\beta=19$, $\mu>0$ we find:
\begin{equation*}
  v^q = 
  \begin{pmatrix}
    3.396\times 10^{-5} & 6.606\times 10^{-4} & 2.419\times 10^{-2} \cr
    3.396\times 10^{-5} & 6.606\times 10^{-4} & 2.419\times 10^{-2} \cr
    3.396\times 10^{-5} & 6.605\times 10^{-4} & 2.419\times 10^{-2}
  \end{pmatrix},
\end{equation*}
\begin{equation*}
  v^\ell = 
  \begin{pmatrix}
    1.023\times 10^{-5} & 2.046\times 10^{-3} & 3.879\times 10^{-2} \cr
    1.023\times 10^{-5} & 2.046\times 10^{-3} & 3.879\times 10^{-2} \cr
    1.023\times 10^{-5} & 2.046\times 10^{-3} & 3.879\times 10^{-2}
  \end{pmatrix}.
\end{equation*}
%

\subsection{The phenomenological neutrino mass matrix}

The constraints on the various products of coupling constants
$\Lambda_i\Lambda_{i'}$, $\lambda_{ijk} \lambda_{i'kj}$, and
$\lambda'_{ijk} \lambda'_{i'kj}$ or their values can be find by a
comparison of theoretical neutrino mass matrix calculated within RpV
MSSM with the phenomenological neutrino mass matrix, derived from
analysis of neutrino data by making viable assumptions.  It is usually
done by assuming that different contributions to theoretical neutrino
mass matrix do not significantly compensate each other, i.e., it is
possible to extract limits on individual contributions without knowing
the others.

The neutrino phenomenological mass matrix in flavor basis, ${\cal M}^{ph}$,
can be written as 
\begin{equation} 
  {\cal M}^{ph} = U^*\, M^{diag}\, U^\dagger,
\end{equation}
where 
\begin{equation}
  M^{diag} \, \equiv \, diag(m_1,\, m_2,\, m_3). 
\end{equation}
Here $m_i$ are the moduli of neutrino mass eigenvalues. The three
neutrino mixing scheme is assumed, namely
\begin{equation}
\nu_{\alpha L}\, =\, U_{\alpha i}\,\nu_{i L}, ~~~~~
\alpha~=~e,\mu,\tau, ~~~~i~=~1,2,3. 
\end{equation} 
$\nu_{\alpha L}$ are flavor neutrino states and $\nu_i$ are the mass 
eigenstates.  

For massive Majorana neutrinos the unitary 
Pontecorvo-Maki-Nakagawa-Sakata (PMNS) matrix $U$ 
in the standard parameterization has the form
%
\begin{widetext}
\begin{eqnarray}
 & & U = 
\left (
    \begin{array}{ccc}
      c_{12} c_{13} & s_{12} c_{13} & s_{13} e^{-i \delta} \\
      -s_{12} c_{23} - c_{12} s_{23} s_{13}e^{i \delta} & c_{12} c_{23} - s_{12}
      s_{23} s_{13}e^{i \delta} & s_{23} c_{13} \\
      s_{12} s_{23} - c_{12} c_{23} s_{13}e^{i \delta} & -c_{12} s_{23} - s_{12}
      c_{23} s_{13}e^{i \delta} & c_{23} c_{13}
    \end{array}
  \right ) 
   \left (
     \begin{array}{ccc}
       1 & 0 & 0 \\
       0 & e^{i \alpha_{12}} & 0 \\
       0 & 0 & e^{i \alpha_{13}} 
     \end{array}
   \right ),
\label{U}
\end{eqnarray}
\end{widetext}
%
where $s_{ij} \equiv \sin\theta_{ij}$, $c_{ij} \equiv \cos\theta_{ij}$.
Three mixing angles $\theta_{ij}$ ($i < j$) vary between 0 and $\pi/2$. 
The $\delta$ is the CP violating Dirac phase and $\alpha_{12}$, $\alpha_{13}$ are
CP violating Majorana phases. Their values vary between $0$ and $2\pi$.
The expression for ${\cal M}^{ph}_{\alpha\beta}$ in terms of $m_i$,
$\theta_{ij}$, $\delta$, $\alpha_{12}$, $\alpha_{13}$ is given in the Appendix.  

The analysis of the Super-Kamiokande atmospheric neutrino data \cite{SK}
and the global analysis of the data of the solar neutrino experiments
and KamLAND experiment \cite{Kamland} yield the following best fit
values of the relevant neutrino oscillation parameters:
\begin{eqnarray}
\Delta m^2_{23} &=& 2.1 \times 10^{-3}~\eV^2,~~~~~~\sin^2 2\theta_{23}~=~1.00,
\nonumber\\
 \Delta m^2_{12}&=& 7.1 \times 10^{-5}~\eV^2,~~~~~~\tan^2\theta_{12}~=~0.40.
\nonumber\\
\label{bestfit}
\end{eqnarray} 
The neutrino mass-squared difference is determined as $\Delta
m^2_{ik}~=~m^2_k-m^2_i$. For the angle $\theta_{13}$ only the upper
bound is known. From exclusion plot obtained from the data of the
reactor experiment CHOOZ \cite{CHOOZ} we have
\begin{equation}
 \sin^2\theta_{13}\, \le \, 5\times 10^{-2}   ~~~~~~(90\%\ \rm{c.l.}).
\label{theta13}
\end{equation} 
The CP violating phases $\delta$, $\alpha_{12}$ and $\alpha_{13}$ remain undetermined.

Neutrino oscillation data are insensitive to the absolute scale of
neutrino masses. The values of neutrino masses depend on the lightest
neutrino mass, on the neutrino mass spectrum and the neutrino
mass-squared differences $\Delta m^2_{12}$ and $\Delta m^2_{23}$. The
neutrino oscillation data are compatible with two types of neutrino mass
spectra:\\ 
i) The normal hierarchy (NH) mass spectrum, which corresponds to the
case
\begin{equation}
  m_1~\ll~m_2~\ll~m_3.
\end{equation}
ii) The inverted hierarchy (IH) of neutrino masses. It is given by the
condition
\begin{equation}
  m_3~\ll~m_1~<~m_2.
\end{equation}

The $0\nu\beta\beta$-decay is one of the most sensitive known ways to
probe the absolute values of neutrino masses and the type of the
spectrum. The most stringent lower bound on the half-life of
$0\nu\beta\beta$-decay were obtained in the Heidelberg-Moscow $^{76}$Ge
experiment \cite{H-M} ($T^{0\nu-exp}_{1/2}\ge 1.9\times 10^{25}$ yr).
By assuming the nuclear matrix element of Ref.~\cite{FedorVogel} we end
up with $|m_{\beta\beta}| =U^{2}_{e1}\,m_{1} + U^{2}_{e2}\,m_{2}
+U^{2}_{e3}\,m_{3} \le 0.55 \eV$, where $U$ is the neutrino mixing
matrix Eq.~(\ref{U}). With this input limit we can find the maximal
allowed values for the matrix elements ${\cal M}^{ph}_{ij}$ of the
neutrino mass matrix, which are as follows:
\begin{eqnarray}
|{\cal M}^{ph-HM}| \le 
  \left (
  \begin{array}{ccc}
    0.55 & 1.29 & 1.29 \\
    1.29 & 1.35 & 1.04 \\
    1.29 & 1.04 & 1.35 
  \end{array}
  \right )~ \eV.
\end{eqnarray}
In the calculation we used best fit values of mass-squared differences
$\Delta m^2_{12}$, $\Delta m^2_{23}$ and mixing angles $\theta_{12}$,
$\theta_{23}$ \cite{SK,Kamland}: $\sin^2\theta_{13}= [0, 5\times
10^{-2}]$, $\delta,~\alpha_{12},\alpha_{13}~=~[0,2\pi]$ and the whole allowed mass
parameter space of neutrinos was taken into account.

The phenomenological neutrino mass matrix can be constructed from the
assumption of normal or inverted hierarchy of neutrino masses. If mass
squared-differences and mixing angles of (\ref{bestfit}) and
(\ref{theta13}) are considered we obtain
\begin{eqnarray*}
  && |{\cal M}^{ph-IH}| = 10^{-2}~\eV \\
  &&\times \left (
  \begin{array}{ccc}
     (1.80,4.52) &  (0.025,3.12) & (0.025,3.12) \\
    (0.025,3.12) &  (0.028,2.40) & (0.945,2.39) \\
    (0.025,3.12) &  (0.945,2.39) & (0.028,2.40) 
  \end{array}
  \right ),
\end{eqnarray*}
\begin{eqnarray*}
  && |{\cal M}^{ph-NH}| = 10^{-4}~\eV \\
  &&\times \left (
  \begin{array}{ccc}
    (0.53,45.2) & (26.9,98.9) & (26.9,98.9) \\
    (26.9,98.9) & (173,254) & (182,254) \\
    (26.9,98.9) & (182,254) & (173,254)
  \end{array}
  \right ).
\end{eqnarray*}
Here, we assumed that the mass of the lightest neutrino is negligible.
The allowed ranges have origin in the uncertainty coming from the
$\theta_{13}$ parameter and the CP violating phases.

\subsection{The magnetic moments of neutrinos}

Once neutrinos are massive, they can have magnetic moments. 
We distinguish magnetic moments of  Dirac and Majorana neutrinos: \\
i) {\it The Dirac magnetic moment}, which connects left-handed electroweak
doublet neutrino $\nu_{iL}$ to a right-handed electroweak singlet neutrino
$\nu_{jR}$ ($i,j = e,~\mu,~\tau$). 
One can express the effective Hamiltonian, which conserves the total lepton
number, as 
\begin{equation}
  H^{D}_{\rm eff} = \frac12\mu^D_{\nu_{ij}} \bar\nu_{iL} 
  \sigma^{\alpha\beta}\nu_{jR} F_{\beta\alpha} + {\rm h.c.},
\label{HeffD}
\end{equation}
where $\mu_{\nu^{D}_{ij}}$ is Dirac diagonal ($i=j$) or transition
($i\ne j$) magnetic moment between states $\nu_{iL}$ and $\nu_{jR}$.

A minimal extension of the SM with a right-handed singlet neutrino 
yields a diagonal neutrino magnetic moment of \cite{marci}
\begin{equation}
\mu_\nu = \frac{3}{4\pi^2} \frac{G_F m_e m_\nu}{\sqrt{2}} \mu_B.
\end{equation}
Here, $\mu_B = e/(2 m_e)$ is the Bohr magneton and $m_\nu$ is the
neutrino mass. By using the neutrino oscillation data one finds $\mu_\nu
\ge 4\times 10^{-20}~\mu_B$. It is believed that larger values of
magnetic moment are possible in extensions of the SM, e.g., in models
with large extra dimensions \cite{edmoh}.  A neutrino magnetic moment
$\mu_\nu$ of the order of $10^{-11}~\mu_B$ might be an explanation of
solar neutrino problem by flipping $\nu_e$ to sterile neutrinos
\cite{volo,barb,grimus,senja,jvns,bala}.

ii) {\it The Majorana magnetic moment} acts between neutrino fields of
the same chirality, namely $\nu_{i L}$ and $\nu_{j L}^c$ As a
consequence it violates the total lepton number by two units ($\Delta L
=2$).  The effective Hamiltonian takes the form
\begin{equation}
  H^M_{\rm eff} = \frac{1}{2} \mu^M_{ij} \bar\nu_{iL} \sigma^{\alpha\beta}
  \nu_{j L}^c F_{\beta\alpha} + {\rm h.c.}
\label{Heff}
\end{equation}
Majorana neutrinos cannot possess a flavour diagonal ($i=j$) magnetic
moment due to the CPT theorem. They can have only transition ($i\ne j$)
magnetic moment \cite{mmmn}.  From the definition one sees that
$\mu_{\nu_{ij}}$ vanishes for $i=j$ and also that magnetic moment is
antisymmetric in indices $i,j$: $\mu_{\nu_{ij}}=-\mu_{\nu_{ji}}$.

The limits on neutrino transition magnetic moments arise from laboratory
measurements of the $\nu_e-e$ scattering cross section using solar,
atmospheric and terrestrial neutrinos \cite{nmmlim}. These experiments
have put upper bounds on $\mu^D_{\nu_{ij}}$, which are as follows:
\begin{eqnarray}
  \mu^D_{\nu_{e j}} &\le& 0.9\times 10^{-10}~\mu_B,\nonumber\\
  \mu^D_{\nu_{\mu j}} &\le& 6.8\times 10^{-10}~\mu_B,\nonumber\\
  \mu^D_{\nu_{\tau j}} &\le& 3.9\times 10^{-7}~\mu_B. 
\end{eqnarray} 
The most restrictive limits remains those from the atrophysical analysis
of the energy-loss rate of stars due to the $\gamma \rightarrow
\nu\overline\nu$ process, which are \cite{raff}
\begin{equation}
\mu^D_{\nu_{i j}} \le 3\times 10^{-12}~\mu_B.
\end{equation} 
The assumption is that this energy loss mechanism of giant red stars
cannot exceed energy loss via weak processes.  For Majorana transition
magnetic moments $\mu^M_{\nu_{ij}}$ the above limits are twice stronger
\cite{sacha}.

The cases of Dirac and Majorana magnetic moments have fundamentally
different physical applications and need to be considered separately.
In this paper we assume massive neutrinos to be Majorana particles.  In
addition, we shall take advantage of the fact that neutrino magnetic
moments have an intimate connection to their masses, which is model
dependent. Recently, the contribution of transition magnetic moments to
the neutrino mass matrix was estimated in \cite{sacha}.  It was found
that if neutrino transition magnetic moments are of the order of their
current upper bound, their contribution to ${\cal M}_{ij}$ can exceed
the experimental value of ${\cal M}^{ph}_{ij}$. The purpose of this
paper is to discuss the impact of the ${\cal M}^{ph}_{ij}$ on the
transition magnetic moments of Majorana neutrinos within the GUT
constrained MSSM with explicit $R$-parity violation.

In order to generate the magnetic moments within the $R$-parity breaking
SUSY one needs to attach a photon to one of the internal lines in the
loops from Figs.~\ref{Fig1} and \ref{Fig2}. For each Feynman diagram we
have two possibilities. One can attach the photon to quark (lepton) or
squark (slepton) internal line of the loop. However, the contributions
coming from photon--sparticle interactions are suppressed by big masses
of the latter and we may safely ignore them.

The contribution coming from the squark--quark loop takes again into
account squark and quark mixing. We end up with
\begin{eqnarray}
  \mu_{\nu_{ii'}}^q &=& (1-\delta_{ii'}) \frac{12 Q_d m_e}{16\pi^2}
  \sum_{jkl} \Bigg \{
  \lambda'_{ijk}\lambda'_{i'kl} \sum_a V_{ja} V_{la}
  \frac{w^q_{ak}}{m_{d^a}} \nonumber \\
  &-& \lambda'_{ijk}\lambda'_{i'lj} \sum_a V_{ka} V_{la}
  \frac{w^q_{aj}}{m_{d^a}} \Bigg \} \mu_B,
\label{Tqq}
\end{eqnarray}
where now the loop integral takes a slightly more complicated form
\begin{equation}
  w^q_{jk} = \frac{\sin 2\theta^k}{2} \left (
    \frac{x_2^{jk}\ln x_2^{jk}-x_2^{jk}+1}{(1-x_2^{jk})^2} -
    (x_2 \to x_1) \right ).
\end{equation}
Here $Q_d = 1/3$ is the $d$-quark charge in units of $e$, and $m_e$
denotes the electron mass. We work in mathematical units, ie. we set
$\hbar=1$. We note that the resulting formula is antisymmetric as
expected.

The contribution from the slepton--lepton loop reads
\begin{eqnarray}
  \mu_{\nu_{ii'}}^\ell &=& (1-\delta_{ii'}) \frac{4 Q_e m_e}{16\pi^2} \\
  &\times& \sum_{jk}
  \lambda_{ijk}\lambda_{i'kj} \left(
    \frac{w^\ell_{jk}}{m_{e^j}} - \frac{w^\ell_{kj}}{m_{e^k}} \right ) 
  \mu_B \nonumber
\label{Tll}
\end{eqnarray}
where the loop integral is as previously equal to
\begin{equation}
  w^\ell_{jk} = \frac{\sin 2\phi^k}{2} \left (
    \frac{y_2^{jk}\ln y_2^{jk}-y_2^{jk}+1}{(1-y_2^{jk})^2} -
    (y_2 \to y_1) \right ).
\end{equation}

It is worthwhile to notice that if quark mixing is neglected,
$\sin{2\theta^k}/2 \approx m_{d^k} m_{\tilde d^k}/ (m_{\tilde d_{1}^k}^2
- m_{\tilde d_{2}^k}^2)$ and the masses of squarks $m_{\tilde d_{1}^k}$,
$m_{\tilde d_{2}^k}$ are replaced with their average value $m_{\tilde
  d^k}$, the expression for $\mu_{\nu_{ii'}}^q$ of Ref. \cite{Bhatta} is
reproduced. The same procedure allows to obtain the result of
\cite{Bhatta} also for $\mu_{\nu_{ii'}}^\ell$.  In this case
$\sin{2\phi^k}/2 \approx m_{e^k} m_{\tilde e^k}/ (m_{\tilde e_{1}^k}^2 -
m_{\tilde e_{2}^k}^2)$ and $m_{\tilde e_{1}^k}$, $m_{\tilde e_{2}^k}$ =
$m_{\tilde e^k}$ (the average mass of sleptons) have to be considered.

%
\begin{table*}
  \caption{\label{tab:cc} The SUSY coefficients converting elements of
    neutrino mass mass matrix to transition magnetic moments of Majorana
    neutrinos (see Eqs. (\ref{eq:muq}) and (\ref{eq:mul}) for
    definition). It is assumed $\mu > 0$.} 
  \begin{ruledtabular}
    \begin{tabular}{ccccccc}
      \multicolumn{4}{c}{SUSY input} &
      \multicolumn{3}{c}{The SUSY conversion coefficient} \\
      \cline{1-4} \cline{5-7}
  $A_0$ & $m_0$ & $m_{1/2}$ & $\tan\beta $ & $f^{q}_{\rm SUSY}$  &
  $f^{q-CKM}_{\rm SUSY}$  & $f^\ell_{\rm SUSY}$  \\ 
 $[\GeV]$ & $[\GeV]$ & $[\GeV]$ &  & $[\eV^{-1}]$ & $[\eV^{-1}]$ & $[\eV^{-1}]$ \\ 
      \hline
      100 & 150 & 150 & 5  & $(0.3,1.0)\times 10^{-16}$ & $(2.8,8.8)\times 10^{-17}$ & $(0.5,1.5)\times 10^{-15}$ \\
          &     &     & 19 & $(0.3,1.2)\times 10^{-16}$ & $(2.8,9.8)\times 10^{-17}$ & $(0.5,1.6)\times 10^{-15}$ \\
      500 &1000 &1000 & 5  & $(1.1,2.8)\times 10^{-18}$ & $(1.0,2.4)\times 10^{-18}$ & $(1.7,4.1)\times 10^{-17}$ \\
          &     &    & 19  & $(1.1,3.1)\times 10^{-18}$ & $(1.0,2.7)\times 10^{-18}$ & $(1.7,4.3)\times 10^{-17}$ \\
    \end{tabular}
  \end{ruledtabular}
\end{table*}
%
%
\begin{table*}
  \caption{\label{tab:mm} The Majorana
    neutrino transition magnetic moment $\mu_{\nu_{ij}}$ ($ij =
    e\mu,~e\tau,~\mu\tau$) derived from phenomenological mass matrices (see
    text for details). The upper bounds on magnetic moments are obtained 
    by using the current neutrino oscillation and the
    $0\nu\beta\beta$-decay data. The predictions for $\mu_{\nu_{ij}}$
    were calculated by assuming the inverted hierarchy or the normal hierarchy
    of neutrino masses and that the lightest neutrino is massless.
    The ranges of allowed values correspond to uncertainty in SUSY coefficient
    $f^{q,l}_{SUSY}$ and in CP-violating violating phases. Two different
    GUT scenarios are considered. The remaining SUGRA parameters are
    $\tan\beta=19$ and $\mu>0$. $\mu_B$ is the Bohr magneton.} 
  \begin{ruledtabular}
    \begin{tabular}{ccccccc}
      \multicolumn{3}{c}{SUSY input in GeV} &
      \multicolumn{4}{c}{Transition magnetic moment $\mu_{\nu_{ij}}$ in $\mu_B$} \\
      \cline{1-3} \cline{4-7}
      $A_0$ & $m_0$ & $m_{1/2}$ & $ij$ & $0\nu\beta\beta$ constraints & inverted hierarchy & normal hierarchy \\ 
      \hline
      &&&& \multicolumn{3}{c}{lepton-slepton loop mechanism} \\
      100 & 150 & 150 & $e\mu, 
                        e\tau$ & $\le (0.62, 2.10) \times 10^{-15}$ & $(0.01, 5.00) \times 10^{-17}$ & $(0.13, 1.60) \times 10^{-17}$ \\
          &     &  & $\mu\tau$ & $\le (0.50, 1.70) \times 10^{-15}$ & $(0.45, 3.80) \times 10^{-17}$ & $(0.87, 4.10) \times 10^{-17}$ \\
      500 &1000 &1000 & $e\mu,
                        e\tau$ & $\le (2.20, 5.50) \times 10^{-17}$ & $(0.04, 13.0) \times 10^{-19}$ & $(0.47, 4.20) \times 10^{-19}$ \\
          &     &  & $\mu\tau$ & $\le (1.80, 4.50) \times 10^{-17}$ & $(0.16, 1.00) \times 10^{-18}$ & $(0.32, 1.10) \times 10^{-18}$ \\
      &&&& \multicolumn{3}{c}{quark-squark loop mechanism (without d-quarks mixing)} \\
      100 & 150 & 150 & $e\mu, 
                        e\tau$ & $\le (0.41, 1.50) \times 10^{-16}$ & $(0.08, 36.0) \times 10^{-19}$ & $(0.08, 1.10) \times 10^{-18}$ \\ 
          &     &  & $\mu\tau$ & $\le (0.33, 1.20) \times 10^{-16}$ & $(0.30, 2.80) \times 10^{-18}$ & $(0.58, 3.00) \times 10^{-18}$ \\ 
      500 &1000 &1000 & $e\mu, 
                        e\tau$ & $\le (1.40, 4.00) \times 10^{-18}$ & $(0.03, 9.60) \times 10^{-20}$ & $(0.30, 3.00) \times 10^{-20}$ \\ 
          &     &  & $\mu\tau$ & $\le (1.20, 3.20) \times 10^{-18}$ & $(1.00, 7.30) \times 10^{-20}$ & $(2.00, 7.80) \times 10^{-20}$ \\
      &&&& \multicolumn{3}{c}{quark-squark loop mechanism (with d-quarks mixing)} \\
      100 & 150 & 150 & $e\mu, 
                        e\tau$ & $\le (0.36, 1.30) \times 10^{-16}$ & $(0.07, 31.0) \times 10^{-19}$ & $(0.76, 9.70) \times 10^{-19}$ \\
          &     &  & $\mu\tau$ & $\le (0.29, 1.00) \times 10^{-16}$ & $(0.27, 2.30) \times 10^{-18}$ & $(0.51, 2.50) \times 10^{-18}$ \\
      500 &1000 &1000 & $e\mu,
                        e\tau$ & $\le (1.30, 3.40) \times 10^{-18}$ & $(0.03, 8.30) \times 10^{-20}$ & $(0.28, 2.60) \times 10^{-20}$ \\  
          &     &  & $\mu\tau$ & $\le (1.10, 2.80) \times 10^{-18}$ & $(0.97, 6.40) \times 10^{-20}$ & $(1.90, 6.80) \times 10^{-20}$ 
    \end{tabular}
  \end{ruledtabular}
\end{table*}
%
%
%
\begin{figure*}[t]
 \includegraphics[width=0.8\textwidth]{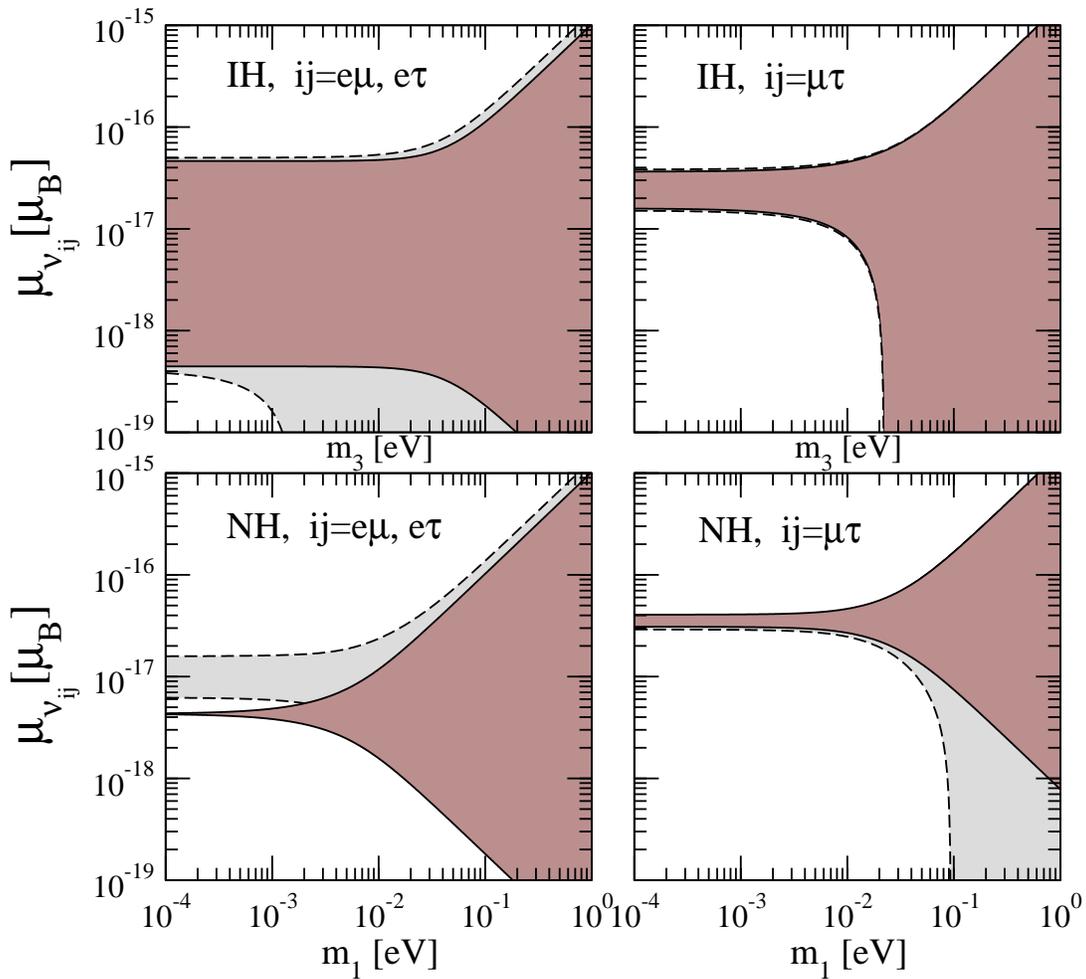}
 \caption{\label{Fig3} Transition magnetic moments of Majorana neutrinos
   $\mu_{\nu_{e\mu}}$, $\mu_{\nu_{e\tau}}$ and $\mu_{\nu_{\mu\tau}}$ as
   function of the lightest neutrino mass $m_3$ (the inverted hierarchy
   of neutrino masses, upper panels) and $m_1$ (the normal hierarchy of
   neutrino masses, lower panels).  The results are presented for
   $\sin^2\theta_{13}=0$ (the region with solid line boundaries) and
   $\sin^2\theta_{13}= 5\times 10^{-2}$ (the region with dashed line
   boundaries).  In the calculation the best fit values of mass-squared
   differences $\Delta m^2_{12}$, $\Delta m^2_{23}$ and mixing angles
   $\theta_{12}$, $\theta_{23}$ were considered \cite{SK,Kamland}. For
   CP-violating phase we assumed $\delta,~\alpha_{12},\alpha_{13}~=~[0,2\pi]$.
   The largest value of the SUSY coefficient $f^{q,l}_{SUSY}$ for the
   set of SUSY parameters $(A_0=100 \GeV$, $m_0=m_{1/2}=150 \GeV)$,
   $\tan\beta=19$ and $\mu > 0$ was taken into account.  IH and NH
   denote the inverted hierarchy and the normal hierarchy of neutrino
   masses, respectively. $\mu_B$ is the Bohr magneton.  }
\end{figure*}
%

The matrices $w^q_{jk}$ and $w^\ell_{jk}$ in flavor space are evaluated
for two representative sets of input parapeters $A_0$, $m_0$, $m_{1/2}$,
$\tan\beta$ and $\mu>0$. By assuming $A_0=100\GeV$, $m_0=m_{1/2}=150
\GeV$, $\tan\beta=19$, $\mu>0$ we end up with
\begin{equation*}
  w^q = 
  \begin{pmatrix}
    2.582\times 10^{-12} & 5.024\times 10^{-11} & 2.332\times 10^{-9} \cr
    6.759\times 10^{-10} & 1.315\times 10^{-8}  & 6.068\times 10^{-7} \cr
    2.750\times 10^{-7}  & 5.350\times 10^{-6}  & 2.421\times 10^{-4}
  \end{pmatrix},
\end{equation*}
\begin{equation*}
  w^\ell = 
  \begin{pmatrix}
    1.223\times 10^{-14} & 2.446\times 10^{-12} & 5.163\times 10^{-11} \cr
    2.691\times 10^{-10} & 5.384\times 10^{-8}  & 1.132\times 10^{-6} \cr
    4.832\times 10^{-8}  & 9.668\times 10^{-6}  & 2.019\times 10^{-4}
  \end{pmatrix}.
\end{equation*}
If larger SUSY mass scale is assumed with $A_0=500 \GeV$, $m_0=m_{1/2}=1000
\GeV$, $\tan\beta=19$, $\mu>0$, we obtain
\begin{equation*}
  w^q = 
  \begin{pmatrix}
    9.895\times 10^{-15} & 1.925\times 10^{-13} & 8.809\times 10^{-12} \cr
    2.780\times 10^{-12} & 5.408\times 10^{-11} & 2.466\times 10^{-9} \cr
    1.383\times 10^{-9}  & 2.690\times 10^{-8}  & 1.216\times 10^{-6}
  \end{pmatrix},
\end{equation*}
\begin{equation*}
  w^\ell = 
  \begin{pmatrix}
    5.316\times 10^{-17} & 1.064\times 10^{-14} & 2.126\times 10^{-13} \cr
    1.301\times 10^{-12} & 2.602\times 10^{-10} & 5.195\times 10^{-9} \cr
    2.755\times 10^{-10} & 5.511\times 10^{-8}  & 1.098\times 10^{-6}
  \end{pmatrix}.
\end{equation*}
%

\section{Results}

The main purpose of our present work is to calculate the transition
magnetic moments for Majorana neutrinos. To achieve this goal one
proceeds as follows. First one finds the values of the loop integrals
$v^\ell$, $v^q$, $w^\ell$ and $w^q$ within some GUT scenario. We have
chosen two sets of parameters: $(A_0=100 \GeV, m_0=m_{1/2}=150 \GeV)$
and $(A_0=500 \GeV, m_0=m_{1/2}=1000 \GeV)$, in both of them keeping
$\tan\beta=19$ and positive $\mu$. This allows us to construct the
theoretical mass matrices, Eqs.~(\ref{Mqq}) and (\ref{Mll}), and
confront them with phenomenological ones. Next we calculate the
contributions from the $\ell$--loop and the $q$--loop. The crucial point
is that we consider each mechanism separately, which means that only one
element from the sums in Eqs.~(\ref{Mqq}) and (\ref{Mll}) is picked up
at a time. Such approach is justified by the usual assumption that there
is no fine-tuning between different contributions, which therefore may
by analyzed separately. Explicitly, the simplified expressions read:
\begin{eqnarray}
  \label{eq:muq}
  \mu_{\nu_{ii'}}^q &\simeq& (1-\delta_{ii'}) \frac43 \mu_B m_{e^1} 
  {\cal M}^q_{ii'}
  \left[ \frac
    {\displaystyle{\sum_a V_{ja} V_{la} w^q_{ak} / m_{q^a}}}
    {\displaystyle{\sum_a V_{ja} V_{la} v^q_{ak} m_{q^a}}}
  \right ]_{\rm max} \nonumber \\
  &\equiv& (1-\delta_{ii'}) {\cal M}^q_{ii'} f^q_{\rm SUSY},
\end{eqnarray}
\begin{eqnarray}
  \label{eq:mul}
  \mu_{\nu_{ii'}}^\ell &\simeq& (1-\delta_{ii'}) 4 \mu_B m_{e^1} 
  {\cal M}^\ell_{ii'}
  \left[ \frac 
    {\displaystyle{w^\ell_{jk} / m_{e^j}}}
    {\displaystyle{v^\ell_{jk} m_{e^j}}}
  \right ]_{\rm max} \nonumber \\
  &\equiv& (1-\delta_{ii'}) {\cal M}^q_{ii'} f^\ell_{\rm SUSY},
\end{eqnarray}
where the $[...]_{\rm max}$ symbol denotes that we have checked all the
combinations of indices $\{j,k,l\}$ in the first case and $\{j,k\}$ in
the second one, and picked up the dominant mechanism.

The coefficients $f^{\ell,q}_{\rm SUSY}$ convert the neutrino masses
into magnetic moments. We list their values, obtained for four different
sets of GUT parameters, in Tab. \ref{tab:cc}. The $f^{q-CKM}_{\rm SUSY}$
parameter takes into account quark mixing, which is absent in $f^q_{\rm
  SUSY}$. This additional mechanism slightly lowers the maximal allowed
value of the magnetic moment One sees also that the dependence on the
$\tan\beta$ parameter is very weak, therefore in the following we will
discuss only the case of large $\tan\beta=19$. For completeness we
include also in the table the lower bounds on the $f$ coefficients,
which were obtained by finding the smallest of the expressions in square
brackets in Eqs. (\ref{eq:muq}) and (\ref{eq:mul}).

The resulting values of $\mu_{\nu_{ii'}}$ for six different scenarios
are presented in Tab.~\ref{tab:mm}. Due to antisymmetricity the diagonal
elements $\mu_{\nu_{ii}}$ are zero. The ranges of values in the first
column are the highest upper limit and lowest upper limit on the
transition magnetic moment, coming from different mechanisms (different
combination of coupling constants $\lambda\lambda$ or
$\lambda'\lambda'$). They were calculated using the ${\cal M}^{ph-HM}$
mass matrix. One clearly sees that in this case the different mechanisms
give comparable results. However, the imposed GUT constraints have a
much bigger impact and introduce two orders of magnitude differences.
The values from the first column for the first set of GUT parameters
($A_0=100\GeV$, $m_0=m_{1/2}=150\GeV$) may be roughly compared with the
predictions published in Ref.~\cite{Bhatta} showing that for this
special case we our results are compatible with the previously
published.

The two remaining columns present the ranges of allowed values of the
magnetic moments if one assumes normal or inverted mass hierarchy of the
neutrinos. These results take into account various mechanisms of
generating $\mu_{\nu_{ii'}}$ as well as the experimental uncertainties
shown in mass matrices ${\cal M}^{ph-IH}$ and ${\cal M}^{ph-NH}$. In
this case the ranges span roughly over 1--2 orders of magnitude. It is
visible that, regardless of the GUT parameters, one should not expect
the magnetic moment to be greater than $10^{-17}\mu_B$, which is not
possible to detect experimentally nowadays.

It is interesting to investigate the influence of quark mixing on the
results. When quark mixing is switched off, the maximal allowed
$\mu^q_{\nu}$ values drop down approximately by 10 to 20\%.

In the case when the lightest neutrino mass is not set to zero, one may
express all the remaining mass eigenstates in terms of $m_{1(3)}$ and
appropiate differences of masses squared, which are known from the
results of the neutrino oscillation experiments. On Fig.~\ref{Fig3} we
present the allowed values for different elements of the neutrino mass
matrix as the function of $m_{1(3)}$ in the case of NH and IH. The
shaded regions correspond to all possible combinations of the two
Majorana $\alpha_{12}$ and $\alpha_{13}$, and one Dirac phase $\delta$,
for $\sin^2 \theta_{13}=0$ (dark shading) and $\sin^2 \theta_{13} \le
0.05$ (lighter shading). One sees clearly, that the phase factors as
well as possible non-zero mass of lightest neutrino modify the results
in a non-trivial fashion.

\section{Summary}

Since the neutrinos have no electric charge they can be Majorana
fermions. A Majorana neutrino is more natural than a Dirac neutrino as
it allows to explain the smallness of their masses in theories with
grand unification. Majorana neutrinos have zero diagonal magnetic
moments while they may in general have transition magnetic moments
$\mu_{\nu_{ij}}$ ($i\ne j$). The question is, how large they are? In
this paper we studied this issue within the GUT constrained MSSM with
explicit $R$-parity violation.

The transition magnetic moments were induced by attaching photons to the
internal lines of lepton-slepton and quark-squarks loop diagrams that
generate neutrino masses. $\mu_{\nu_{ij}}$ has been written as a product
of SUSY conversion coefficients $f_{SUSY}$, which reflects the
dependence on SUSY parameter space and quark mixing, and the element of
Majorana neutrino mass matrix in flavor basis.

Two possible GUT scenarios were considered: $A_0=100\GeV$,
$m_0=m_{1/2}=150\GeV$, and $A_0=500\GeV$, $m_0=m_{1/2}=1000\GeV$, in
both of them having $\tan\beta=19$ and positive $\mu$ parameter.  Our
study showed that $f_{SUSY}$ is nearly insensitive on the value of
$\tan\beta$ and that its dependence on the scale of GUT common mass and
coupling constants parameters $m_0$, $m_{1/2}$, and $A_0$ dominates. The
difference in $f_{SUSY}$ for these two scenarios reach around two orders
of magnitude. It is worth mentioning that contributions to
$\mu_{\nu_{ij}}$ from lepton-slepton mechanisms dominates over
contributions from quark-squark diagrams.

The transition magnetic moments were calculated by considering three
differently constructed phenomenological neutrino mass matrices. The
elements of the first one (${\cal M}^{ph-HM}$) represent the maximal
allowed values, if neutrino oscillation data \cite{SK,Kamland} and the
lower bound on the $0\nu\beta\beta$-decay half-life of $^{76}$Ge
\cite{H-M} are taken into account. The corresponding upper limits on
$\mu_{\nu_{e\mu}}$, $\mu_{\nu_{e\tau}}$ and $\mu_{\nu_{\mu\tau}}$ have
been found to be of the order of $10^{-15}~\mu_B$ and $10^{-17}~\mu_B$
for the assumed two SUSY scenarios.

Two phenomenological neutrino mass matrices were generated by
considering normal and inverted hierarchy of neutrino masses and that
the mass of the lightest neutrino is zero. This allowed to make
predictions for transition magnetic moments of Majorana neutrinos.  For
low (large) SUSY parameter scale we found the maximal value $5\times
10^{-17}~\mu_B$ ($1.3\times 10^{-18}~\mu_B$). The dependence of the
results on the neutrino mass pattern is weak.

The magnetic moments $\mu_{\nu_{e\mu}}$, $\mu_{\nu_{e\tau}}$ and
$\mu_{\nu_{\mu\tau}}$ were calculated also as a function of the lightest
neutrino mass. The best fit values of neutrino oscillation parameters
were considered. Only for the mixing angle $\theta_{13}$ it was assumed
$\sin^2\theta_{13}=0$ and $\sin^2\theta_{13}= 5\times 10^{-2}$.  It is
worth mentioning that the maximal mixing of atmospheric neutrinos
($\theta_{23}=\pi/4$) implies that the allowed ranges of
$\mu_{\nu_{e\mu}}$ and $\mu_{\nu_{e\tau}}$ are the same, if the Dirac
and Majorana CP violating phases vary from 0 to $\pi/2$. The results
show that especially $\mu_{\nu_{e\mu, e\tau}}$ (inverted hierachy of
neutrino masses) depends strongly on the values of CP phases.

The obtained results show that the values of the Majorana transition
magnetic moments might be significantly above the scale of the
Dirac-type magnetic moment in minimal extension of the SM with
right-handed neutrinos. However, the maximal values are still too small
to be tested in the present laboratory experiments or to have
astrophysical consequences.

{\bf Acknowledgments} This work was supported by the VEGA Grant agency
of the Slovak Republic under contract No.~1/3039/06, by the EU ILIAS
project under contract RII3-CT-2004-506222, by the DFG project 436 SLK
17/298 and the Polish State Committee for Scientific Research under
grants no. 2P03B~071~25 and 1P03B~098~27. One of us (MG) greatly
acknowledges the financial support from the Foundation for Polish
Science.

\section{Appendix}

In this appendix, for covenience of the reader we present explicit
expressions for the neutrino mass matrix in flavor basis, ${\cal
  M}_{\alpha\beta}$ ($\alpha,\beta=e,\mu,\tau$), as functions of the
moduli of neutrino mass eigenvalues $m_i$, of mixing angles
$\theta_{ij}$ and of the CP violating phases $\delta$, $\alpha_{12}$,
$\alpha_{13}$ \cite{frigerio}.  The matrix element ${\cal M}$ is symmetric,
i.e., it is defined by six elements:
%
\begin{widetext}
\begin{eqnarray}
{\cal M}_{ee} &=&
~~c_{13}^{2}\,c_{12}^{2}\,{m_1}\,+\,
c_{13}^{2}\,s_{12}^{2}\,{m_2}\,{e^{-i2\alpha_{_{12}}}}\,+\,
s_{13}^{2}\,{e^{2\,i\delta}}\,{m_3}\,{e^{-i2\alpha_{_{13}}}},
\nonumber\\
\nonumber \\
{\cal M}_{e\mu} &=&
-\,c_{12}\,c_{13}\, 
\left( \,c_{23}\,s_{12}\,+\,c_{12}\,s_{23}\,s_{13}\,{e^{-i\delta}}\,
\right) {\it {m_1}}
\nonumber\\
&&+\,c_{13}\,s_{12}
\left( \,c_{23}\,c_{12}\,-\,s_{23}\,s_{12}\,s_{13}\,{e^{-i\delta}}
\right)\,{\it {m_2}} \,{e^{-i 2\alpha_{_{12}}}} 
\,+\,c_{13}\,s_{23}\,s_{13}\,{e^{i\delta}}\,{m_3}\,{e^{-i2\alpha_{_{13}}}},
\nonumber\\
\nonumber\\
{\cal M}_{e\tau} &=&
-\,c_{12}\,c_{13}\,
\left(-\,s_{23}\,s_{12}\,+\,c_{23}\,c_{12}\,s_{13}\,{e^{-i\delta}}
\right)\,{m_1}
\nonumber\\
&&-\,c_{13}\,s_{12}\,
\left(\,c_{12}\,s_{23}\,+\,c_{23}\,s_{12}\,s_{13}\,{e^{-i\delta}}\,
\right)
{m_2}{e^{-i2\alpha_{_{12}}}}\,+\,
c_{23}\,c_{13}\,s_{13}\,{e^{i\delta}}\,{m_3}\,{e^{-i2\alpha_{_{13}}}},
\nonumber\\
\nonumber\\
{\cal M}_{\mu\mu} &=&
~~\left(
c_{23}^2\,s_{12}^2\,+\,
2\,c_{23}\,c_{12}\,s_{23}\,s_{12}\,s_{13}\,{e^{-i\delta}}\,+
c_{12}^2\,s_{23}^2\,s_{13}^2\,{e^{- 2 i \,\delta}}
\right)\,m_1\,
\nonumber\\
&&+\left(
c_{23}^2\,c_{12}^2\,-\,
2\,c_{23}\,c_{12}\,s_{23}\,s_{12}\,s_{13}\,{e^{-i\delta}}\,+\,
s_{23}^2\,s_{12}^2\,s_{13}^2\,{e^{- 2 i \,\delta}}
\right) 
\,{m_2}\,{e^{-i2\alpha_{_{12}}}} 
\,+\,
c_{13}^2\,
s_{23}^2\,
{m_3}\,{e^{-i2\alpha_{_{13}}}},
\nonumber\\
\nonumber\\
{\cal M}_{\mu\tau} &=&
-\left( 
c_{23}\,s_{23}\,s_{12}^{2}\,
-\,c_{23}^{2}\,c_{12}\,s_{12}\,s_{13}\,{e^{-i\delta}}\,
+\,c_{12}\,s_{23}^{2}\,s_{12}\,s_{13}\,{e^{-i\delta}}\,
-\,c_{23}\,c_{12}^{2}\,s_{23}\,s_{13}^{2}\,{e^{-2\,i\delta}}\,
\right) {\it {m_1}}
\nonumber\\
&&- \left( 
c_{23}\,c_{12}^{2}\,s_{23}\,
+\,c_{23}^{2}\,c_{12}\,s_{12}\,s_{13}\,{e^{-i\delta}}\,
-\,c_{12}\,s_{23}^{2}\,s_{12}\,s_{13}\,{e^{-i\delta}}\,
-\,c_{23}\,s_{23}\,s_{12}^{2}\,s_{13}^{2}\,{e^{-2\,i\delta}}\,
\right) \,{m_2}\,{e^{-i2\alpha_{_{12}}}}
\nonumber\\
&&\,+\,c_{23}\,c_{13}^{2}\,s_{23}\,{m_3}\,{e^{-i2\alpha_{_{13}}}},
\nonumber\\
\nonumber\\
{\cal M}_{\tau\tau} &=&
~~\left( s_{23}^{2}\,s_{12}^{2}
\,-\,2\,c_{23}\,c_{12}\,s_{23}\,s_{12}\,s_{13}\,{e^{-i\delta}}
\,+\,c_{23}^{2}\,c_{12}^{2}\,s_{13}^{2}\,{e^{-2\,i\delta}}\right) {m_1}
\nonumber\\
&&+\left( 
c_{12}^{2}\,s_{23}^{2}\,+
\,2\,c_{23}\,c_{12}\,s_{23}\,s_{12}\,s_{13}\,{e^{-i\delta}}
+c_{23}^{2}\,s_{12}^{2}\,s_{13}^{2}\,{e^{-2\,i\delta}}
\right) 
\,{m_2}\,{e^{-i2\alpha_{_{12}}}}\,
+\,c_{23}^{2}\,c_{13}^{2}\,{m_3}\,{e^{-i2\alpha_{_{13}}}}.
\nonumber\\
\end{eqnarray}
\end{widetext}
%

The maximal mixing of atmospheric neutrinos ($\theta_{23}=\pi/4$)
implies symmetry among some of elements of the neutrino mass matrix
${\cal M}_{\alpha\beta}$.  If CP-violating phases are considered as free
parameters the allowed ranges of $|{\cal M}_{e\mu}|$ and $|{\cal
  M}_{e\tau}|$ ($|{\cal M}_{\mu\mu}|$ and $|{\cal M}_{\tau\tau}|$)
coincide each with other.  If additional restriction is taken into
account, namely $\sin^2(\theta_{13})=0$, we obtain
\begin{eqnarray}
{\cal M}_{ee} &=&
~~c_{12}^{2}\,{m_1}\,+\,s_{12}^{2}\,{m_2}\,{e^{-i2\alpha_{_{12}}}}
\nonumber\\
{\cal M}_{e\mu} &=& \frac{1}{\sqrt{2}} 
c_{12}\,s_{12}\,\left(-\,{\it {m_1}}+{\it {m_2}} \,{e^{-i 2\alpha_{_{12}}}} 
\right),
\nonumber\\
{\cal M}_{e\tau} &=&
-\frac{1}{\sqrt{2}} 
c_{12}\,s_{12}\,\left(-\,{\it {m_1}}+{\it {m_2}} \,{e^{-i 2\alpha_{_{12}}}} 
\right),
\nonumber\\
{\cal M}_{\mu\mu} &=&
~~\frac{1}{2}\left(\,s_{12}^2\,\,m_1\,+\,c_{12}^2\,\,m_2\,{e^{-i2\alpha_{_{12}}}}
\,+\,m_3\,{e^{-i2\alpha_{_{13}}}}\right),  
\nonumber\\
{\cal M}_{\mu\tau} &=&
~~-\frac{1}{2}\left(\, s_{12}^2\,\,m_1\,+\,c_{12}^2\,\,m_2\,{e^{-i2\alpha_{_{12}}}}
\,-\,m_3\,{e^{-i2\alpha_{_{13}}}}\right),  
\nonumber\\
{\cal M}_{\tau\tau} &=&
~~\frac{1}{2}\left(\,s_{12}^2\,\,m_1\,+\,c_{12}^2\,\,m_2\,{e^{-i2\alpha_{_{12}}}}
\,+\,m_3\,{e^{-i2\alpha_{_{13}}}}\right).
\nonumber\\  
\end{eqnarray}
In this case $|{\cal M}_{e\mu}|=|{\cal M}_{e\tau}|$ and ${\cal
  M}_{\mu\mu}={\cal M}_{\tau\tau}$. In addition, for $\delta$,
$\alpha_{12}$, $\alpha_{13}$ between $0$ and $2\pi$ the maximal and minimal
absolute values of all three elements ${\cal M}_{\mu\tau}$, ${\cal
  M}_{\mu\mu}$ and ${\cal M}_{\tau\tau}$ are the same.



\end{document}